\documentstyle[aps,preprint,tighten]{revtex}
\def\beq{\begin{equation}}
\def\eeq{\end{equation}}
\def\beqa{\begin{eqnarray}}
\def\eeqa{\end{eqnarray}}

\def\GeV{\nobreak\,\mbox{GeV}}
\def\fm{\nobreak\,\mbox{fm}}
\def\mb{\nobreak\,\mbox{mb}}

\begin{document}
\title{\sc Charmonium-Hadron Cross Section in a Nonperturbative QCD Approach} 
\author{H.G. Dosch$^1$\thanks{e-mail: H.G.Dosch@ThPhys.Uni-Heidelberg.DE}, \ 
F.S. Navarra$^{2}$\thanks{e-mail: navarra@if.usp.br}, \ 
M. Nielsen$^{2}$\thanks{e-mail: mnielsen@if.usp.br} \ and \
M. Rueter$^3$\thanks{e-mail: rueter@post.tau.ac.il}\\[0.3cm]
{\it $^1$Institut f\"ur Theoretische Physik, Universit\"at Heidelberg}\\
{\it Philosophenweg 16, D-6900 Heidelberg, Germany}\\[0.1cm]
{\it $^2$Instituto de F\'{\i}sica, Universidade de S\~{a}o Paulo} \\
{\it C.P. 66318, 05315-970  S\~{a}o Paulo, SP, Brazil}\\[0.1cm]
{\it $^3$Department of High Energy Physics, Tel-Aviv University}\\
{\it 69978 Tel-Aviv, Israel}}
\maketitle
\vspace{1cm}
\begin{abstract}
We calculate the nonperturbative $J/\Psi-N$ and $\Psi'-N$ cross sections
with the model of the stochastic vacuum  which has been succesfully 
applied in many high energy reactions.  We also give a quantitative discussion 
of pre-resonance formation and medium effects. 

\end{abstract}
\vspace{1cm}

During the next year the first data on heavy ion collisions at high energy
($\sqrt{s}= 200\, \GeV$ per nucleon pair) will be available at RHIC. As it is
well known, one of the main goals of this machine is to find and study the 
plasma of quarks and gluons (QGP) \cite{QM97,QM99}. 
The search for this state of matter has 
started in the early eighties and since then has been subject of intense 
debate \cite{QM97,QM99}. Recent results on Pb-Pb collisions, taken at lower 
energies at CERN-SPS, have attracted a great attention and increased the 
hope that a new phase of nuclear matter is ``just around the corner''.  

The signature of QGP formation has been and still remains a theoretical 
and experimental challenge. Indeed there is so far no ``crucial test'' 
able to disentangle the possible new phase from  the dense hadronic 
background. Among the proposed signatures the most interesting is the 
suppression of  $J/\Psi$ \cite{matsatz}. 
This suppression was observed experimentally by the  
NA38 collaboration in 1987 in collisions with light ions and also and more 
dramatically by the NA50 collaboration in 1995-1996 in Pb-Pb collisions
\cite{na38,na50}.

Whereas the old data could be reasonably well explained by a
``conventional'' approach the new  Pb-Pb results 
of the NA50 collaboration created a big controversy \cite{kharz}. 
The entire set of the 
$J/\Psi$ data from pA and AB collisions available before the advent of the 
Pb beam at CERN SPS has been found to be consistent with the nuclear 
absorption model. However, in the case of the new Pb-Pb data, the density 
of secondaries, which (together with primary nucleons flying around) are 
presumably responsible for the charmonium absorption, is so high that the  
hadronic system in question is hardly in a hadronic phase.  Nevertheless, 
a conventional treatment of the problem is not yet discarded \cite{spi}. 

Reliable values for 
the 
charmonium nucleon cross sections are of crucial importance in the present
context. One needs to know the 
cross section $\sigma_{J/\psi}$ in order 
to predict a nuclear suppression  of $J/\Psi$ without assuming a so called
``deconfining regime''. Estimates using perturbative QCD 
give values which are too
small to explain the observed absorption conventionally, but they are 
certainly not reliable for that genuine nonperturbative problem. A 
nonperturbative estimate may be
tried by applying vector dominance to 
$J/\Psi$  and  $\Psi'$ photoproduction.  
In this way a 
cross section of $\sigma_{J/\psi} \simeq 1.3$ mb for 
$\sqrt{s} \simeq 10$ GeV and 
$ \sigma_{\psi'}/\sigma_{J/\psi} \simeq 0.8$ has been obtained
\cite{HK,fs}. 
In ref. \cite{dgkp} strong arguments against the vector dominance with only 
few intermediate vector mesons were put forward even in the case of the 
production of light vector mesons and which apply a fortiori to the 
production of heavy vector mesons. The 
instability of the vector dominance model can be seen from a more refined 
multichannel analysis 
\cite{HK} where a value  $\sigma_{J/\psi} \simeq 3-4$ mb
has been obtained. Even this value  is too small in order to explain the 
absorption in p-A collisions which is of the order
$\sigma_{\psi}^{abs} \simeq 7.3$ mb \cite{klns,satz}

These hadron-hadron 
cross sections involve nonperturbative aspects of QCD dynamics and therefore 
require a nonperturbative model to be calculated. In a  recent letter
\cite{gfssg98} the nonperturbative QCD contribution to the charmonium-nucleon
cross section was evaluated by using an interpolation formula for the 
dependence of the cross section on the transverse size of a quark-gluon
configuration.
In this work we calculate the $J/\Psi$ and $\Psi'$ - nucleon cross sections
in a specific nonperturbative model of QCD: the  model of the stochastic 
vacuum (MSV)
\cite{Dosch:1987,Dosch:1988,Simonov:1988,dosch}. 
It has been applied to a large 
number of hadronic and photoproduction  processes (including photoproduction 
of $J/\Psi$) with remarkably good success. Its application to $J/\Psi$ and 
$\Psi'$  nucleon scatttering
 is straightforward. We also investigate the influence of the nuclear 
matter and arrive at rather stringent limits for the cross sections in 
an environment different from the vacuum and where the properties of the 
medium are reflected in a shift  of the $J/\Psi$ mass \cite{fe}.

The basis of the MSV is the calculation of the scattering
amplitude of two colourless dipoles \cite{dgkp,dosch} based on a semiclassical
treatment developped by Nachtmann \cite{nach}. For details we refer to  
the literature  and show here only some
intermediate steps necessary for  the understanding of the text.  The
dipole-dipole scattering amplitude is  expressed as the expectation value of two
Wegner-Wilson loops with lightlike sides and  transversal extensions $\vec
r_{t 1}$ and $\vec r_{t 2}$ respectively. This leads to a  profile function $J(\vec b,
\vec r_{t 1},\vec r_{t 2})$ from which  hadron-hadron scattering amplitudes are
obtained by integrating over  different dipole sizes with the
transversal densities of the hadrons as weight functions according to 
\begin{equation}
\sigma^{tot}_{J/\Psi} = \int \, d^2 b \, d^2 r_{t 1} \,  d^2 r_{t 2}
 \, 
 \rho_{J/\Psi}(\vec r_{t 1}) \,
\rho_N(\vec r_{t 2}) \,  J(\vec b, \vec r_{t 1},\vec r_{t 2})\; .
\label{totcross}
\end{equation} 
Here $\rho_{J/\Psi}(\vec r_{t 1})$ and $\rho_N(\vec r_{t 2})$ are the transverse 
densities of the $J/\Psi$ and nucleon respectively. 

The basic ingredient of the model is the gauge invariant correlator of
two gluon field strength tensors. The latter is characterized by two
constants: the value at zero distance, the gluon condensate $<g^2FF>$,
and the correlation length $a$. We take these values from previous
applications of the model  \cite{dgkp} ( and literature quoted there):
\beq 
<g^2FF>= 2.49 \rm{ GeV}^4 \qquad a= 0.346 \rm{ fm} \; .
\label{const}
\eeq

The wave functions of the proton have been determined from
proton-proton and proton-antiproton scattering respectively. It turns out that 
the best description for the nucleon transverse density is given by that 
of a quark diquark system with transversal distance ${\vec r}_t$ and density: 
\beq
\rho_N(\vec{r_t})=
|\Psi_p (\vec{r}_t)|^2= \frac{1}{2\pi}\frac{1}{S_p^2} \, e^{-\frac{
|\vec{r}_t|^2}{2S_p^2}}\; .
\label{Gausswfmeson}
\eeq
The value of the extension parameter,  $S_p=0.739\,\fm$, obtained from 
proton-proton scattering  agrees
very well with that obtained from the electromagnetic form factor in a similar 
treatment \cite{Paulus}.

For the wave function of the $J/\Psi$ we used two approaches:

\noindent

\noindent
1) A numerical solution of the Schroedinger equation with the standard 
Cornell potential \cite{cornell}:
\beq
V = - \frac{4}{3} \frac{\alpha_s}{r} + \sigma r\; .
\eeq

2) A Gaussian wave function determined by the electromagnetic decay
width of the $J/\Psi$ which has been used in a previous investigation of
$J/\Psi$ photoproduction \cite{dgkp}. 

For the $\Psi'$ no analysis of photoproduction in the model has been
made so we use only the solution of the Schroedinger equation.

The linear potential can be calculated in the model
of the stochastic vacuum which  yields  the string tension:
\beq 
\sigma={8\kappa\over81\pi}<g^2 FF> a^2\; =0.179 \GeV^2\; ,
\label{sigma}
\eeq
where the parameter $\kappa$  has been detemined in lattice calculations to be 
$\kappa = 0.8 $ \cite{lat}.

The other parameters, the charmed (constituent) mass and the
(frozen) strong coupling can be adjusted in order to give the correct
$J/\Psi$ and $\Psi'$ mass difference and the $J/\psi$ decay width
\beq
m_c= 1.7 \rm{ GeV } \qquad  \alpha_s= 0.39 \; .
\label{malpha} 
\eeq
We also use the  standard Cornell model parameters \cite{cornell}: 
\beq
\alpha_s = 0.39, \sigma = 0.183 \, 
\GeV^2 \mbox{ and }  m_c = 1.84 \, \GeV. \label{cornell} 
\eeq

From the numerical solution $\psi(|\vec r\,|)$ of the Schroedinger equation 
the transversal density is projected:
\beq
\rho_{J/\Psi} (\vec{r}_t) = \int\left|\psi(\sqrt{\vec{r}_t\,^2+ r_3^2})
\right|^2 dr_3\; ,
\eeq
where $\vec{r}_t $ is the $J/\Psi$ transversal radius.

Given the values of $\alpha_s$,   $\sigma$  and  $m_c$  we solve 
the  non-relativistic Schroedinger equation numerically, obtain the wave 
function,  compute the transverse wave function and plugg it into the MSV 
calculation 
\cite{dosch}. The results are shown in Table I. In this table  
$\sqrt{<r^2>}$ is the root of the mean square distance of  quark and antiquark
and  $\sqrt{<r_t^2>}$ is the root of the mean square transversal distance 
of quark and antiquark. Wave function A) is the one obtained with the 
parameters given by  Eqs.~(\ref{sigma}) and (\ref{malpha}). Wave function B) 
corresponds to the standard Cornell model 
parameters \cite{cornell} Eq.~(\ref{cornell}).

\vskip 5mm
\begin{center}
\begin{tabular}{||l|c|c|c||}
\hline
Wave function & $\sqrt{<r^2>}$ fm & $\sqrt{<r_t^2>}$ fm & $\sigma_{tot}$ [mb]\\
\hline\hline
$J/\Psi(1S)$&&&\\
A &0.393&0.321&4.48\\
B  &0.375 &0.306 &4.06\\
C &  &  & 4.69\\ \hline
$\Psi(2S)$&&&\\
A:&0.788&0.640&17.9\\
\hline 
\end{tabular}
\end{center}
\begin{center}
\bf{TABLE I} {\small $J/\Psi-N$ and $\Psi'-N$ cross section. A and B: 
numerical 
solution of the Schroedinger equation with parameters in 
Eqs.~(\protect\ref{malpha}) 
and (\protect\ref{cornell}) respectively. C: Cross section obtained by
the weighted average of the longitudinally and transversely polarized 
$J/\Psi$ wave functions of ref.~\protect\cite{dgkp}.}
\end{center}
\vskip5mm

In ref.~\cite{dgkp}
a gaussian ansatz was made to construct vector meson wave functions that
describe well the electromagnetic decay of the vector meson and photo
and electroproduction cross sections. In Table I,  wave function C) gives 
the result for the $J/\Psi-N$ cross section obtained with the 
weighted average of the longitudinally and transversely polarized 
$J/\Psi$ wave functions
from \cite{dgkp} with transversal sizes $\sqrt{<r_t^2>}=0.327\;\fm$ and
0.466 fm.

Averaging over our results for different wave functions,  
our final result for the $J/\Psi-N$ cross section is 
\beq
\sigma_{J/\psi}=4.4\pm0.6 \mb\;.
\label{resu}
\eeq

The error is an estimate of uncertainties coming from the wave function and
the model. The only other nonperturbative calculation of the
$J/\Psi-N$ cross section that we are aware of was done in ref.\cite{gfssg98}
and the obtained cross section was  $\sigma_{J/\psi}=3.6$ mb, in a fair
agreement with our result and with recent analysis of $J/\Psi$
photoproduction data \cite{HK}. For $\Psi'$ our cross section is also of
the same order as the value obtained  in \cite{gfssg98}:
$\sigma_{\psi'} =  20.0$ mb. 

Since one of the possible explanations of the observed $J/\Psi$ suppression
is based on the pre-resonance absorption model \cite{klns} 
we present numerical calculations of the 
nucleon - pre-resonant charmonium state cross section,  $\sigma_{\psi}$.  
In the pre-resonance absorption model, the pre-resonant 
charmonium state is either interpreted as a color-octet, $(c \overline c)_8$,  and 
a gluon in the hybrid  $(c \overline c)_8 - g$ state, or as a coherent  
$J/\Psi - \Psi'$ mixture. We use a gaussian transverse wave function,  
as in Eq.~(\ref{Gausswfmeson}),  to represent a state 
with transversal radius 
$\sqrt{<r_t^2>} \simeq 0.82 \,\sqrt{<r^2>} \, = \sqrt{2} S_{\psi}$ ($ S_{\psi}$ 
is the pre-resonance extension parameter analogous to $S_{p}$).  
With the knowledge of the wave functions and transformation properties of 
the constituents  we can compute the total cross 
section given by the MSV. The 
resulting nucleon - pre-resonant charmonium state cross section
 will be different if the pre-resonant charmonium state
consists of entities in the adjoint representation 
(as $(c \bar c)_8 - g$) or in the fundamental representation
(as a $J/\Psi - \Psi'$ mixture), the relation being $\sigma_{\rm adjoint}
=\frac{2 N_C^2}{N_C^2-1}\sigma_{\rm fundamental}$, with $N_C=3$. 
In Table II we show the results for these two 
possibilities and different values of the transverse radius. 

\vspace{.5cm}
\begin{center}
\begin{tabular}{||c|c|c||}  \hline
$\sqrt{<r_T^2>}$ & $\sigma_{c \bar c}$ &$\sigma_{(c \bar c)_8-g}$  \\
(fm) & (mb) & (mb) \\
\hline\hline
0.20 & 1.79 &4.02 \\
\hline
0.25 & 2.76 &6.21\\
\hline
0.30 & 3.96 &8.91\\
\hline
0.35 & 5.30 &11.92\\
\hline
0.40 & 6.81 &15.32\\
\hline
0.45 & 8.50 &19.12\\
\hline
0.50 & 10.28 &23.13\\
\hline
\end{tabular}
\end{center}
\begin{center}
\bf{TABLE II} {\small The cross section charmonium-nucleon for gaussian 
wave-functions and different values of the  transverse
radius ($\sqrt{<r_{t}^2>}$) of  the $c \bar c$ 
in a singlet state (first row) or in a
hybrid $(c \bar c)_8 - g$ state (second row).}
\end{center}
\vspace{.5cm}

From our results we can see that a cross-section $\sigma_{\psi}^{abs} 
\simeq 6-7$ mb, needed to explain the $J/\Psi$ and $\Psi'$ suppression in
p-A collisions in the pre-resonance absorption model \cite{klns,spi}, 
is consistent with a pre-resonant charmonium state of 
size $\simeq 0.50-0.55$ fm if it is a $J/\Psi - \Psi'$ mixture 
or $\simeq 0.30-0.35$ fm for a $(c \bar c)_8 - g$ state.

So far the calculations were done with the vacuum values of the
correlation length and gluon condensate generally used in the MSV \cite{dosch}.
However, since the interaction between the charmonium and the nucleon
occurs in a hadronic medium, these values may change. Indeed,
lattice calculations \cite{lat,miller} show that both
the correlation length and the gluon condensate tend to decrease in a dense 
(or hot) medium. The reduction of the string tension, $\sigma$,
leads to two competing effects,  which can be quantitatively compared in the
MSV. On one hand the cross section  tends to decrease strongly when the gluon
condensate or the correlation length  decrease. On the other hand, when the
string tension is reduced the  $c -  \overline c$ state becomes less confined
and will have a larger radius,  which, in turn, would lead to a larger cross
section for interactions with the  nucleons in the medium. It is of major
interest to  determine which of these  effects is dominant.

The dependence of the total cross section, Eq.~ (\ref{totcross}), on the 
extension parameters $S_p$ and $S_\psi$ is quite well
parametrized \cite{thesis} as:
\beq
\sigma_{J/\psi} \propto <g^2FF>^2
  a^{10} \left({S_p\over a}\right)^{1.5}
\left({S_\psi\over a}\right)^2 
\eeq
In the MSV the string tension, $\sigma$, is related to the gluon condensate
and to the correlation length through Eq.~(\ref{sigma}).
Therefore, the dependence of the cross section on the string tension
and correlation length is approximately given by:
\beq
\sigma_{\psi} \propto \sigma^2 a^6 \left({S_p\over a}\right)^{1.5}
\left({S_\psi\over a}\right)^2 \; .
\eeq
In a rough approximation the hadron radii can be estimated, using the Ritz 
variational principle, to be:
\beq
S\propto\left({1\over \sigma}\right)^{1/3}\; ,
\eeq
and thus we finally obtain the following three possibilities to express the 
cross section as a function of the string tension, $\sigma$, the correlation 
length, $a$, and the gluon condensate, $<g^2FF>$:
\beq
\sigma_{\psi\,N} \propto \left\{\begin{array}{c}
\sigma^{5/6} a^{5/2}\\
 \sigma^{25/12} <g^2FF>^{-5/4}\\
 <g^2FF>^{5/6} a^{25/6}\end{array}\right. 
\; 
\label{fi}
\eeq

From the equation above we see that the final effect of the medium is a 
reduction in the cross section. We can also see that   
a 10\% variation in the parameters lead to large variations on the 
cross sections. 
Using the values  of the correlation length and the gluon condensate reduced
by 10\%: $a=0.31$ fm , $\langle g^2 FF\rangle=2.25$ GeV$^4$, we obtain a
40\% reduction in the cross sections. Taking this reduction into account 
the cross sections obtained in this work are smaller than the ones needed
(both in Refs. \cite{klns} and \cite{spi}) to explain experimental data.
However, since in our model the nucleon - pre-resonant charmonium state 
cross section is much bigger than the $J/\Psi - N$ cross section if the 
pre-resonance charmonium state is a hybrid  $(c \overline c)_8 - g$ state, the
reduction in the cross sections due to medium effects favors the pre-resonant
model for the hadronic explanation of the observed $J/\Psi$ suppression.

In order to get more precise results we have varied only one  the parameters
$a$ and $<g_s^2 FF>$ and kept the other fixed. This was done in such a way as
to decrease the string tension according to equation (5) to the values given
in  Table III, first row. The numerically evaluated values for the mass of
the $J/\Psi$ (serving as a physical measure of the change) and the cross
sections are given in rows 2 to 4, using Gaussian wave functions determined by
the variational principle. 

\vskip5mm
\begin{center}
\begin{tabular}{||l|c|c|c||}\hline
string tension [GeV$^2$] & $\Delta E$ [MeV] & 
$\sigma_{tot}$ [mb]&$\sigma_{tot}$ [mb]\\
(GeV$^2$) & MeV & (mb)& (mb) \\ 
 & &$a$ const.&$<g^2 FF>$ const.\\
\hline\hline
$\sigma_0$ & 0 & 4.48& 4.48\\
$0.9 \sigma_0$& -33 & 3.86&3.38\\
$0.75 \sigma_0$ & - 83 & 3.27&2.2\\
$0.50 \sigma_0$& -174&2.0&0.83\\
$0.25 \sigma_0$ & -275 & 0.91&0.13\\
\hline
\end{tabular}
\end{center}
\begin{center}
\bf{TABLE III} {\small $J/\Psi$-Nucleon total cross sections as function of 
the string tension with either the correlation length  or the gluon condensate
kept constant. $\Delta E$ is the mass decrease of the $J/\Psi$  due to 
the change of string tension.}
\end{center}
\vskip5mm


To summarize, we calculated the nonperturbative $J/\Psi-N$ and $\Psi'-N$ 
cross sections with the MSV. The basic ingredient of the model is the gauge
invariant correlator of two gluon field strength tensors which is 
characterized by two constants: the gluon condensate and the gluon field
correlation length. Using for these quantities values fixed in previous 
applications and using well accepted charmonium wave functions we obtain
$\sigma_{J/\psi\, N}\sim4$ mb and $\sigma_{\psi'\,N}\sim 18$ mb. An interesting
prediction of the MSV is the strong depedence on the parameters of the QCD
vacuum which will most likely lead to a drastic reduction of them at higher
temperatures and perhaps also at higher densities.

\vskip5mm

\underline{Acknowledgements}: This work has been supported by FAPESP (under
project \#: 98/2249-4)
and CNPq.  We would like to warmly thank E. Ferreira for 
instructive discussions in the early stage of this work.
F.S.N. and M.N. would like to thank the Institut f\"ur Theoretische 
Physik at the University of Heidelberg for its hospitality 
during their stay in Heidelberg.

\vspace{1cm}

\end{document}